# A Novel Approach for Complete Identification of Dynamic Fractional Order Systems Using Stochastic Optimization Algorithms and Fractional Calculus

Deepyaman Maiti, Mithun Chakraborty, and Amit Konar

Department of Electronics and Telecommunication Engineering, Jadavpur University
Kolkata - 700032, West Bengal, India
E-mail: deepyamanmaiti@gmail.com, mithun.chakra108@gmail.com, and konaramit@yahoo.co.in

**Abstract** - This contribution deals with identification of fractional-order dynamical systems. System identification, which refers to estimation of process parameters, is a necessity in control theory. Real processes are usually of fractional order as opposed to the ideal integral order models. A simple and elegant scheme of estimating the parameters for such a fractional order process is proposed. This method employs fractional calculus theory to find equations relating the parameters that are to be estimated, and then estimates the process parameters after solving the simultaneous equations. The said simultaneous equations are generated and updated using particle swarm optimization (PSO) technique, the fitness function being the sum of squared deviations from the actual set of observations. The data used for the calculations are intentionally corrupted to simulate real-life conditions. Results show that the proposed scheme offers a very high degree of accuracy even for erroneous data.

## I. Introduction

Proper estimation of the parameters of a real process, fractional or otherwise, is a challenge to be encountered in the context of system identification [1] - [3]. Accurate knowledge of the parameters of a system is often the first step in designing controllers. Many statistical and geometric methods such as least square and regression models are widely used for real-time parameter estimation. The problem of parameter estimation becomes more difficult for a fractional order system compared to an integral order one. The real world objects or processes that we want to estimate are generally of fractional order [4]. A typical example of a non-integer (fractional) order system is the voltage-current relation of a semi-infinite lossy RC line or diffusion of heat into a semi-infinite solid, where heat flow $q(t)$ is equal to the half-derivative of temperature $T(t)$: $\frac{d^{0.5}T(t)}{dt^{0.5}} = q(t)$.

So far, however, the usual practice when dealing with a fractional order process has been to use an integer order approximation. In general, this approximation can cause significant differences between a real system and its mathematical model. Disregarding the fractional order of the system was caused mainly by the non-existence of simple mathematical tools for the description of such systems. Since major advances have been made in this area recently, it is possible to consider also the real order of the dynamical systems. Such models are more adequate for the description of dynamical systems with distributed parameters than integer-order models with concentrated parameters. With regard to this, in the task of identification, it is necessary to consider also the fractional-order of the dynamical system.

Most classical identification methods cannot cope with fractional order transfer functions. Yet, this challenge must be overcome if we want to design a proper adaptive or self-tuning fractional order controller. Need for design of adaptive controllers gives an impetus to finding accurate schemes for system identification.

Computation of transfer characteristics of the fractional order dynamic systems has been the subject of several publications [5] – [8], e.g. by numerical methods [7], as well as by analytical methods [8]. In this paper we propose a method for parameter identification of a fractional order system for a chosen structure of the model using fractional calculus theory to obtain simultaneous equations relating the unknown parameters and then solving these equations to obtain accurate estimates. This method enables us to work with the actual fractional order process rather than an integer order approximation. Using it in a system with known parameters will do the verification of the correctness of the identification.

We first consider that the fractional powers are constant and display the accuracy of the proposed method both when random corruptions are absent and present. Then we remove this limitation and propose a scheme using the particle swarm optimization (PSO) algorithm by which a fractional order system can be completely identified with a high degree of accuracy even in presence of significant quantities of error in the readings. PSO, a stochastic optimization strategy from the family of evolutionary computation, is a biologically-inspired technique originally proposed by Kennedy and Eberhart in [12]. The PSO algorithm will generate process models guided by the fractional differintegral definitions and optimize the models after comparing the simulated outputs from such models with the set of outputs obtained from the actual fractional order system (for the same input).

Section II discusses the basics of fractional calculus and PSO.

## II. Fractional Calculus and PSO Algorithm

Sub-section A gives the necessary theory and formulae of fractional differintegral. The relevant facts about the PSO algorithm are mentioned in sub-section B.

### A. Theory of Fractional Calculus

The fractional calculus is a generalization of integration and derivation to non-integer order operators. At first, we generalize the differential and integral operators into one fundamental operator ${}_aD_t^\alpha$ where: ${}_aD_t^\alpha = \frac{d^\alpha}{dt^\alpha}$ for $\Re(\alpha) > 0$; 1 for $\Re(\alpha) = 0$; $\int_a^t (d\tau)^{-\alpha}$ for $\Re(\alpha) > 0$.

The two definitions used for fractional differintegral are the Riemann-Liouville definition [9], [10] and the Grunwald-Letnikov definition [9], [11].

The Grunwald-Letnikov definition is

$${}_aD_t^\alpha f(t) = \lim_{h \to 0} \frac{1}{h^\alpha} \sum_{j=0}^{\left[\frac{t-a}{h}\right]} (-1)^j \binom{\alpha}{j} f(t - jh) \quad (1)$$

($[y]$ means the greatest integer not exceeding y).

Derived from the Grunwald-Letnikov definition, the numerical calculation formula of fractional derivative can be achieved as

$${}_{t-L}D_t^\alpha f(t) \approx h^{-\alpha} \sum_{j=0}^{[L/T]} b_j f(t - jh) \quad (2)$$

L is the length of memory. T, the sampling time always replaces the time increment h during approximation. The weighting coefficients $b_j$ can be calculated recursively by:

$$b_0 = 1, b_j = \left(1 - \frac{1+\alpha}{j}\right) b_{j-1}, (j \geq 1). \quad (3)$$

### B. Particle Swarm Optimization (PSO)

The PSO algorithm [12] - [14] attempts to mimic the natural process of group communication of individual knowledge, which occurs when a social swarm elements flock, migrate, forage, etc. in order to achieve some optimum property such as configuration or location.

The 'swarm' is initialized with a population of random solutions. Each particle in the swarm is a different possible set of the unknown parameters to be optimized. Representing a point in the solution space, each particle adjusts its flying toward a potential area according to its own flying experience and shares social information among particles. The goal is to efficiently search the solution space by swarming the particles toward the best fitting solution encountered in previous iterations with the intent of encountering better solutions through the course of the process and eventually converging on a single minimum error solution.

Let the swarm consist of N particles moving around in a D-dimensional search space. Each particle is initialized with a random position and a random velocity. Each particle modifies its flying based on its own and companions' experience at every iteration. The $i^{th}$ particle is denoted by $X_i$, where $X_i = (x_{i1}, x_{i2}, \ldots, x_{iD})$. Its best previous solution (pbest) is represented as $P_i = (p_{i1}, p_{i2}, \ldots, p_{iD})$. Current velocity (position changing rate) is described by $V_i$, where $V_i = (v_{i1}, v_{i2}, \ldots, v_{iD})$. Finally, the best solution achieved so far by the whole swarm (gbest) is represented as $P_g = (p_{g1}, p_{g2}, \ldots, p_{gD})$.

At each time step, each particle moves towards pbest and gbest locations. The fitness function evaluates the performance of particles to determine whether the best fitting solution is achieved. The particles are manipulated according to the following equations:

$$v_{id}(t+1) = \omega v_{id}(t) + c_1 \varphi_1 (p_{id}(t) - x_{id}(t)) + c_2 \varphi_2 (p_{gd}(t) - x_{id}(t))$$

$$x_{id}(t+1) = x_{id}(t) + v_{id}(t+1).$$

(The equations are presented for the $d^{th}$ dimension of the position and velocity of the $i^{th}$ particle.)

Here, $c_1$ and $c_2$ are two positive constants, called cognitive learning rate and social learning rate respectively, $\varphi_1$ and $\varphi_2$ are two random functions in the range [0,1], $\omega$ is the time-decreasing inertia factor designed by Eberhart and Shi [9]. The inertia factor balances the global wide-range exploitation and the nearby exploration abilities of the swarm.

## III. Process of Identification for Constant Fractional Powers

We have considered a fractional order process whose transfer function is of the form $\frac{1}{a_1 s^\alpha + a_2 s^\beta + a_3}$. The orders of fractionality $\alpha$ and $\beta$ are known and the coefficients $a_1$, $a_2$ and $a_3$ are to be estimated. One important advantage of the proposed scheme is that we do not require to know the ranges of variation of $a_1$, $a_2$ and $a_3$. It should be noted that without loss of generality, we may presume the dc gain to be unity so that the dc gain and its possible fluctuations are included in the coefficients $a_1$, $a_2$ and $a_3$. If C(s) is the output and R(s) the input,

$$\frac{C(s)}{R(s)} = \frac{1}{a_1 s^\alpha + a_2 s^\beta + a_3},$$

$$\Rightarrow R(s) = a_1 s^\alpha C(s) + a_2 s^\beta C(s) + a_3 C(s)$$

In time domain,

$$r(t) = a_1 D^\alpha c(t) + a_2 D^\beta c(t) + a_3 c(t) \quad (4)$$

$$\Rightarrow r(t) \approx a_1 T^{-\alpha} \sum_{j=0}^{[L/T]} b_j c(t - jT) + a_2 T^{-\beta} \sum_{j=0}^{[L/T]} b_j c(t - jT) + a_3 c(t)$$

The proposed scheme requires sampled input at time instant t and sampled outputs at time instants t, t – T, t – 2T, t – 3T,.......... Sampled outputs are required for a time length L previous to t, T being the sampling time. Calculation of fractional derivatives and integrals requires the past history of the process to be remembered. So more the value of L, the better.

Thus the values of $D^\alpha c(t)$ and $D^\beta c(t)$ can be calculated using (1), (2) and (3) so that (4) reduces to the form $a_1 p + a_2 q + a_3 r = s$, where $p, q, r, s$ are constants whose values have been determined.

Let us assume that we have a set of sampled outputs c(t) from the system for unit step test signal.
That is, we have

$$u(t) = a_1 D^\alpha c(t) + a_2 D^\beta c(t) + a_3 c(t). \qquad (5)$$

Now there are three unknown parameters, namely $a_1$, $a_2$ and $a_3$. So we need three simultaneous equations to solve for them. One equation is (5). We will integrate both sides of (5) to obtain

$$\int u(t)dt = \int [a_1 D^\alpha c(t) + a_2 D^\beta c(t) + a_3 c(t)]dt.$$

This gives us

$$r(t) = a_1 D^{\alpha-1} c(t) + a_2 D^{\beta-1} c(t) + a_3 D^{-1} c(t) \qquad (6)$$

Here r(t) signifies unit ramp input and c(t) is the output due to unit step input. Thus we have derived a second equation relating $a_1$, $a_2$ and $a_3$.
The third equation will be obtained by integrating both sides of (6). This gives us

$$p(t) = a_1 D^{\alpha-2} c(t) + a_2 D^{\beta-2} c(t) + a_3 D^{-2} c(t) \qquad (7)$$

Here p(t) signifies parabolic input and c(t) is the output due to unit step input.
It can be seen that (5), (6), (7) are distinct equations in $a_1$, $a_2$ and $a_3$. So we can solve them simultaneously to identify the three unknown parameters $a_1$, $a_2$ and $a_3$.
As we have displayed elsewhere, direct application of the above scheme gives very satisfactory results when the readings c(t) are accurate. If we now add an error component e(t) to c(t) to have a distorted output waveform $c(t) \equiv c(t) + e(t)$ from which we want to make our identification, (5) will be transformed to

$$u(t) = a_1 D^\alpha [c(t) + e(t)] + a_2 D^\beta [c(t) + e(t)] + a_3 [c(t) + e(t)] \qquad (8)$$

So (8) will not give an accurate relation between $a_1$, $a_2$ and $a_3$ due to the presence of the terms $a_1 D^\alpha e(t)$, $a_2 D^\beta e(t)$ and $a_3 e(t)$. Likewise, equations obtained by applying the transformation $c(t) \equiv c(t) + e(t)$ on (6) and (7) will also be inaccurate. Our aim will be to minimize this inaccuracy.
One significant fact we observed is that for the same random error waveform e(t), $D^{\alpha_1} e(t) << D^{\alpha_2} e(t)$ if $\alpha_1 < 0$ and $\alpha_2 > 0$ when in effect, $D^{\alpha_1} e(t)$ becomes an integration.
A rigorous mathematical proof explaining this observation cannot be presented here for constraint of space. But for now, a philosophical explanation may be put forward as follows. The physical significance of differentiation is the slope of the function we want to differentiate (although this is not strictly the case for fractional differentiation), whereas integration deals with the area under the curve. The random error component we considered consists of fluctuations having both positive and negative values. Thus an integration operation over this error waveform can be expected to yield quite a low value, since the areas with opposing signs should nullify the effects of each other. On the other hand, integration of the output waveform will give a high positive result, since the output waveform can assume non-negative values only. So the effect due to the error component is minimized.
However, we cannot say anything definite about the results of differentiation operations, in fact no pattern can be ascribed to the result obtained by differentiating either the output or the error waveforms.
To support our contention, in table 1, we tabulate the values of $D^\alpha e(t)$ for 10 different sets of randomly generated e(t) with $\alpha$ = 1.5, 1.2, 0.9, 0.6, 0.3, -0.3, -0.6, -0.9, -1.2, -1.5. The amplitude of e(t) varies between –0.01 and 0.01. Length of memory = 10 seconds, i.e. the fractional derivatives are calculated at time t = 10 seconds. Sampling rate is once in 0.001 seconds.

The transfer function of our system is $\dfrac{1}{a_1 s^\alpha + a_2 s^\beta + a_3}$, and as we are well aware, $\alpha, \beta > 0$ for a practical system, so that in (8), the orders of derivation $\alpha, \beta$ are positive. To remedy this, we can perform a simple transformation on the transfer function of the system, which we can write as $\dfrac{s^{-n}}{a_1 s^{\alpha-n} + a_2 s^{\beta-n} + a_3 s^{-n}}$, where $(n-1) < \alpha < n$ and $\alpha > \beta$.

Table 1. Variation of $D^\alpha e(t)$ with α. (The 10 sequences e(t) are consecutive and independent.)

| e(t) | $D^\alpha e(t)$ for derivation order α | | | | | | | | | |
|---|---|---|---|---|---|---|---|---|---|---|
| | α = 1.5 | α = 1.2 | α = 0.9 | α = 0.6 | α = 0.3 | α = -0.3 | α = -0.6 | α = -0.9 | α = -1.2 | α = -1.5 |
| 1 | -435.7842 | -50.7575 | -5.7583 | -0.6287 | -0.0661 | -0.0008 | 0.0001 | 0.0006 | 0.0013 | 0.0025 |
| 2 | -603.6659 | -59.5517 | -5.7933 | -0.5742 | -0.0617 | -0.0013 | -0.0005 | -0.0004 | -0.0002 | 0.0002 |
| 3 | 424.4136 | 44.8209 | 4.7948 | 0.5242 | 0.0581 | 0.0002 | -0.0002 | -0.0001 | -0.0001 | -0.0003 |
| 4 | -256.3730 | -26.5634 | -3.0495 | -0.3928 | -0.0549 | -0.0011 | -0.0002 | -0.0001 | -0.0002 | -0.0005 |
| 5 | -107.8138 | -12.0636 | -1.4119 | -0.1631 | -0.0164 | 0.0004 | 0.0004 | 0.0005 | 0.0008 | 0.0013 |
| 6 | 642.4164 | 78.0679 | 9.1798 | 1.0311 | 0.1069 | 0.0006 | -0.0001 | -0.0002 | -0.0006 | -0.0012 |
| 7 | 184.7026 | 22.6486 | 2.6896 | 0.3051 | 0.0321 | 0.0000 | -0.0001 | -0.0001 | 0.0002 | 0.0003 |
| 8 | -393.9215 | -45.1151 | -5.0296 | -0.5467 | -0.0562 | 0.0001 | 0.0002 | 0.0000 | -0.0001 | -0.0002 |
| 9 | -109.5421 | -4.2756 | 0.4383 | 0.1517 | 0.0272 | 0.0006 | 0.0001 | 0.0000 | -0.0001 | 0.0000 |
| 10 | -32.4628 | -7.6152 | -1.4990 | -0.2670 | -0.0439 | -0.0008 | 0.0002 | 0.0007 | 0.0014 | 0.0024 |

Proceeding as before we can now obtain our three simultaneous equations as:

$$D^{-n}u(t) = (a_1 D^{\alpha-n} + a_2 D^{\beta-n} + a_3 D^{-n})[c(t)+e(t)] \quad (9)$$
$$D^{-n-1}u(t) = (a_1 D^{\alpha-n-1} + a_2 D^{\beta-n-1} + a_3 D^{-n-1})[c(t)+e(t)] \quad (10)$$
$$D^{-n-2}u(t) = (a_1 D^{\alpha-n-2} + a_2 D^{\beta-n-2} + a_3 D^{-n-2})[c(t)+e(t)] \quad (11)$$

It can now be checked that all orders of derivation are now negative so that we will actually be performing fractional order integrations rather than fractional order differentiations.

## IV. Illustration

Let the process whose parameters are to be estimated is
$$\frac{1}{a_1 s^{2.23} + a_2 s^{0.88} + a_3}.$$

The input considered is $r(t) = 1$ i.e. unit step.
Synthetic data for $c(t)$ is created using $a_1 = 0.8$, $a_2 = 0.5$ and $a_3 = 1$, i.e. the values of $c(t)$ are obtained at different time instants (using a MATLAB program) assuming a process with transfer function $\frac{1}{0.8 s^{2.23} + 0.5 s^{0.88} + 1}$. The simultaneous equations corresponding to (9), (10) and (11) are

$$D^{-3}u(t) = (a_1 D^{-0.77} + a_2 D^{-2.12} + a_3 D^{-3})[c(t)+e(t)]$$
$$D^{-4}u(t) = (a_1 D^{-1.77} + a_2 D^{-3.12} + a_3 D^{-4})[c(t)+e(t)]$$
$$D^{-5}u(t) = (a_1 D^{-2.77} + a_2 D^{-4.12} + a_3 D^{-5})[c(t)+e(t)]$$

Length of memory L = 10 seconds and T = 0.001 seconds is used to calculate the fractional derivatives.

We will display the accuracy of identification when the output readings used to calculate the fractional derivatives are ideal and also when they are erroneous to the extent of a random error component in the range [−0.05, 0.05] in each reading. This error component is quite large since the output response is often below unity. The output response of the system for unit step input is shown both in presence and absence of the error component.

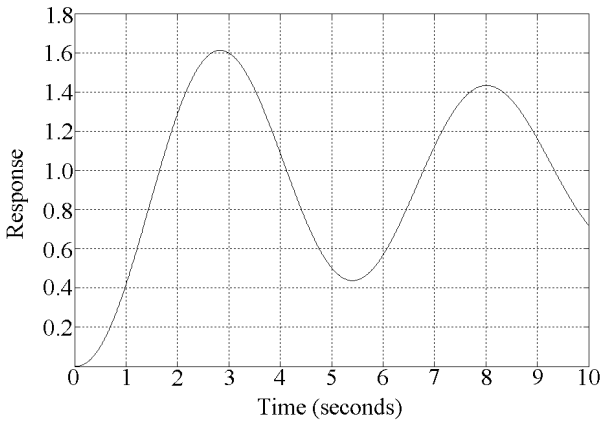

Fig. 1 Actual unit step response.

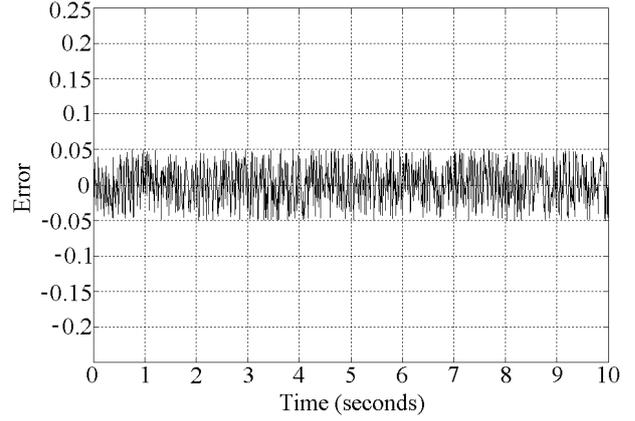

Fig. 2 Random error component added.

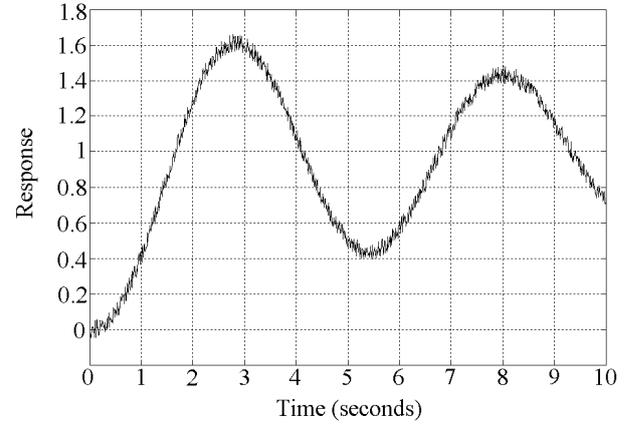

Fig. 3 Randomly corrupted unit step response.

### A. Ideal Case: e(t) = 0 for all t

The following derivatives are then calculated numerically using (1), (2) and (3):

$D^{-0.77}c(t) = 6.1777$, $\quad D^{-2.12}c(t) = 51.3011$,
$D^{-3}c(t) = 136.1477$, $\quad D^{-1.77}c(t) = 32.2818$,
$D^{-3.12}c(t) = 152.6826$, $\quad D^{-4}c(t) = 314.8183$,
$D^{-2.77}c(t) = 108.0207$, $\quad D^{-4.12}c(t) = 342.4005$,
$D^{-5}c(t) = 576.6986$.

The set of simultaneous equations is
$$\begin{bmatrix} 6.1777 & 51.3011 & 136.1477 \\ 32.2818 & 152.6826 & 314.8183 \\ 108.0207 & 342.4005 & 576.6986 \end{bmatrix} \begin{bmatrix} a_1 \\ a_2 \\ a_3 \end{bmatrix} = \begin{bmatrix} 166.7167 \\ 416.9167 \\ 834.1670 \end{bmatrix}$$

After solving, we have $a_1 = 0.8001, a_2 = 0.4996, a_3 = 1.0000$ as the unknown parameters. The errors in estimating them are respectively 0.0125%, 0.0800% and 0%.
Summation of square errors of this process model outputs relative to the output data set for unit step input is 0.0030.

## B. Non-ideal Case: Each Element in e(t) is Between –0.05 and 0.05

To each reading c(t) is added a random error component varying between –0.05 and 0.05.

We proceed as before to obtain the set of simultaneous equations as

$$\begin{bmatrix} 6.1798 & 51.3179 & 136.1948 \\ 32.2919 & 152.7357 & 314.9314 \\ 108.0577 & 342.5242 & 576.9207 \end{bmatrix} \begin{bmatrix} a_1 \\ a_2 \\ a_3 \end{bmatrix} = \begin{bmatrix} 166.7167 \\ 416.9167 \\ 834.1670 \end{bmatrix}$$

After solving, we have $a_1 = 0.7992, a_2 = 0.4996, a_3 = 0.9996$ as the unknown parameters. The errors in estimating them are respectively 0.1000%, 0.0800% and 0.0400%.

Summation of square errors of this process model outputs relative to the actual output data set for unit step input is 0.0062.

So, we can conclude that the proposed identification scheme is highly accurate even in presence of large quantities of random error.

## V. Complete Identification by Applying PSO

From sections III and IV, we understand that if we can find the fractional powers α and β first, we can then accurately identify the other parameters (coefficients). We will employ a standard two-parameter optimization PSO algorithm where the two parameters to be optimized are the fractional powers of the system.

Let the range of variation of α is $\alpha_{min}$ to $\alpha_{max}$ and the range of variation of β is $\beta_{min}$ to $\beta_{max}$. Then, initially a random solution set of $(\alpha, \beta)$ is generated with the position vector limits $\{(\alpha_{min}, \alpha_{max}), (\beta_{min}\beta_{,max})\}$. The position vector $(\alpha, \beta)$ is updated using the PSO dynamics [13]. So the solution space is two-dimensional, as are the position and velocity vectors.

The fitness function $F(\alpha, \beta)$ is derived as follows. We apply a test signal R(s)=1/s (unit step) to the actual system and obtain sampled values of output c(t). For $\alpha = \alpha$ and $\beta = \beta$, the coefficient terms $a_1, a_2, a_3$ is generated. Let, for the process model, the output response for unit step input (obtained by simulation) is p(t). We will define a parameter $F = \sum_{t=i}^{f} [c(t) - p(t)]^2$, which gives a measure of the deviation of the output of the trial process model from the output of the actual process. F is the fitness function that the PSO algorithm will try to minimize. At F = 0, the unknown parameters $(\alpha, \beta)$ are optimized. The corresponding $a_1, a_2, a_3$ are the identified coefficients. The process model corresponding to the optimized solution set should provide output identical to c(t) for unit step input.

Clearly, our only source of information about the actual process is the set of output readings from it. The PSO algorithm will try to find a process model that matches these readings. So we have to perform one transformation from s-domain to discrete time domain, since the actual readings will obviously be in time domain. An alternative approach is to convert all data into z-domain. But then we would have been required to perform two transformations: discrete time domain to z-domain and s-domain to z-domain. Our approach is thus simpler.

Admittedly, the same problem could have been solved by a straightforward application of a five-parameter estimation PSO with the five parameters as $\alpha, \beta, a_1, a_2, a_3$. But by the direct use of fractional calculus definitions we have simplified the problem to the optimization of just two parameters (the fractional powers). The other three parameters can be derived from these two and are not independent. So we have presented a more efficient approach.

## VI. Illustration of Complete Identification

We have used a sampling frequency of 20 hertz, i.e. the output waveform is sampled once every 0.05 seconds.

The range of variation α is 2.0 to 2.4, that of β is 0.7 to 1.1. The PSO parameters used are: the inertia factor ω decreases linearly from 0.9 to 0.4, the cognitive learning rate $c_1$=1.4, the social learning rate $c_2$=1.4.

Number of PSO particles in the population is 10. The PSO algorithm is run for 40 iterations, and this is kept as the stop criterion.

The position vectors of the particles are randomly initialized in the range [(2.0, 2.4), (0.7, 1.1)]. The velocity vectors are randomly initialized in the range [(-0.2, 0.2), (-0.2, 0.2)]. The limits on the position vectors are [(2.0, 2.4), (0.7, 1.1)]. No limit is kept on the velocity vectors.

Now the PSO algorithm is run. After 40 iterations, the best particle has the position (2.2301, 0.8808), which is the identification of the fractional powers. The identification of the three coefficients ($a_1, a_2, a_3$) is (0.7996, 0.4998, 1.0001). The value of the best fitness after 40 iterations is 4.7388 x $10^{-6}$.

So, the identified system is $\dfrac{1}{07996s^{2.2301} + 0.4998s^{0.88} + 1.0001}$. The percentage errors in identification for $\alpha, \beta, a_1, a_2, a_3$ are respectively 0.0045, 0.0909, 0.0500, 0.0400, 0.0100.

Some relevant graphs are presented below.

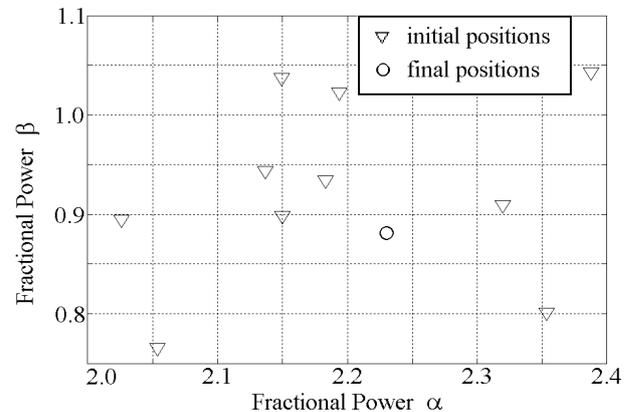

**Fig. 4 Convergence of the PSO particles.**

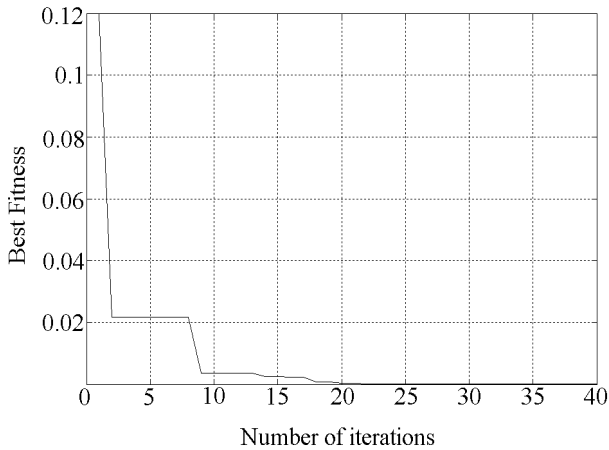

**Fig. 5 Variation of best fitness with number of iterations for one run.**

Results from 10 other sample runs (using identical conditions) are shown in table 2.

**Table 2. Ten sample runs of the identification algorithm.**

| α | β | $a_1$ | $a_2$ | $a_3$ | Fitness X $10^6$ |
|---|---|---|---|---|---|
| 2.2301 | 0.8810 | 0.7996 | 0.4998 | 1.0002 | 3.8420 |
| 2.2304 | 0.8814 | 0.7993 | 0.5000 | 1.0002 | 5.2414 |
| 2.2302 | 0.8812 | 0.7994 | 0.4999 | 1.0002 | 3.9793 |
| 2.2302 | 0.8811 | 0.7995 | 0.4999 | 1.0002 | 3.5780 |
| 2.2301 | 0.8810 | 0.7995 | 0.4998 | 1.0002 | 3.6754 |
| 2.2301 | 0.8808 | 0.7997 | 0.4998 | 1.0001 | 4.7388 |
| 2.2302 | 0.8812 | 0.7994 | 0.4999 | 1.0002 | 3.6306 |
| 2.2301 | 0.8810 | 0.7995 | 0.4998 | 1.0002 | 3.8420 |
| 2.2302 | 0.8811 | 0.7995 | 0.4999 | 1.0002 | 3.5800 |
| 2.2303 | 0.8812 | 0.7994 | 0.4999 | 1.0002 | 3.8707 |

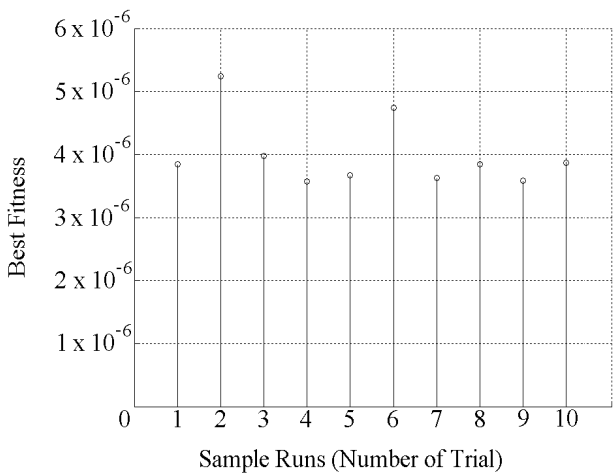

**Fig. 6 Variation of best fitness with sample runs (using data from table 2).**

## VII. Comparison, Comments and Conclusion

An elegant method for the identification of parameters of a fractional order system is proposed. The process of identification can actually be implemented by a simple computer program in C or MATLAB. Of course, the same method can easily be employed to estimate the parameters of an integer order process model as well.

The challenge in fractional order system identification is that the fractional powers are not restricted to assume only discrete integral values, but are distributed in a continuous interval. For two integral order systems, identical time domain responses mean identical transfer functions. But for fractional order systems, we often find that a better identification of the actual process has actually a lower fitness than a worse model.

The method of finding a relation between the coefficients by use of fractional calculus simplifies a complex problem. The subsequent application of an evolutionary algorithm thus provides very accurate estimations. So the proposed scheme is both simple and efficient. Therein lies its merit.